\documentclass[11pt,a4paper]{article}
\pdfoutput=1
\usepackage{jheppub,bm,color,bbold}
\usepackage[T1]{fontenc}

\newcommand{\M}{\hat{M}}
\newcommand{\m}{\hat{m}}
\newcommand{\D}{\mathcal{D}} 

\graphicspath{{./1_Figures/}}

\DeclareMathOperator\diag{diag}
\DeclareMathOperator\ee{e}
\DeclareMathOperator\tr{tr}

\renewcommand{\Re}{\mathrm{Re}}

\renewcommand{\bar}[1]{\overline{#1}}
\newcommand{\calO}{\mathcal{O}}
\newcommand{\dd}{\mathrm{d}}
\newcommand{\bep}{\begin{pmatrix}} 
\newcommand{\eep}{\end{pmatrix}}

\newcommand{\SU}{\text{SU}}

\newcommand{\U}{\text{U}}
\newcommand{\1}{\mathbb{1}}
\newcommand{\RR}{\mathbb{R}}

\renewcommand{\epsilon}{\varepsilon}

\def\ba#1\ea{\begin{align}#1\end{align}}
\def\akakko#1{\left\langle #1 \right\rangle}
\def\mkakko#1{\left( #1 \right)}
\def\ckakko#1{\left\{ #1 \right\}}
\def\kkakko#1{\left[ #1 \right]}

\title{Heavy-tailed chiral random matrix theory}
\author{Takuya Kanazawa}
\affiliation{iTHES Research Group and Quantum Hadron Physics Laboratory, 
RIKEN, Wako, Saitama 351-0198, Japan}
\emailAdd{takuya.kanazawa@riken.jp}
\preprint{RIKEN-QHP-194}

\abstract{ 
We study an unconventional chiral random matrix model with a heavy-tailed 
probabilistic weight. The model is shown to exhibit chiral 
symmetry breaking with no bilinear condensate, in analogy to 
the Stern phase of QCD. We solve the model analytically and 
obtain the microscopic spectral density and the smallest 
eigenvalue distribution for an arbitrary number of flavors and 
arbitrary quark masses. Exotic behaviors such as 
non-decoupling of heavy flavors and a power-law tail of 
the smallest eigenvalue distribution are illustrated. 
}

\allowdisplaybreaks
\begin{document}
\maketitle
\section{Introduction}

Random matrix theory has flourished as a versatile tool in theoretical 
and mathematical sciences 
over decades \cite{Guhr:1997ve,Mehta_book,handbook:2010}.  
In Quantum Chromodynamics (QCD),  the development of chiral random matrix 
theory (chRMT) \cite{Shuryak:1992pi,Verbaarschot:1993pm,Verbaarschot:1994qf} 
(also called the Wishart-Laguerre ensemble) that extends the traditional Wigner-Dyson classes 
has helped us gain a profound understanding of the link between 
dynamical mass generation of fermions and 
spectral statistics of the Dirac operator \cite{Banks:1979yr}.  The 
chRMT has also been used as a simplistic Ginzburg-Landau-type model of 
QCD at finite temperature and density \cite{Stephanov:1996ki,Halasz:1998qr}. 
Exact spectral correlations of a non-Hermitian Dirac operator at nonzero 
chemical potential were also worked out 
\cite{Osborn:2004rf,Akemann:2004dr,Akemann:2005fd,Kanazawa:2009en,Akemann:2010tv}. 
On the practical side, chRMT has enabled accurate determinations of 
low-energy constants in lattice QCD near the chiral limit \cite{Damgaard:2006pu,Fukaya:2007fb}. 
We refer to \cite{Verbaarschot:1997bf,Verbaarschot:2000dy,Akemann:2007rf,
Verbaarschot:2009jz,Kanazawa:2012zzr} for reviews on chRMT. 

In RMT there are various choices for the probabilistic weight of random matrix elements. 
While the independent Gaussian distribution is the simplest from a mathematical point of view, 
it often turns out that distributions that deviate from Gaussian lead to the same spectral 
correlations in the limit of large matrices. 
This robustness of RMT is known as \emph{universality} \cite{DeiftBook:2009}. 
However, when the deformation of the weight is strong enough, results begin to differ from 
those of the Gaussian ensemble. 
Such random matrix ensembles with heavy-tailed weights 
have found applications to disordered conductors and financial statistics, as reviewed in \cite{Burda:review}.  
In general, when the rotational invariance of matrices is broken 
as in the L\'{e}vy matrix ensemble \cite{Cizeau:1994zz}, the models tend 
to be analytically intractable. 
Alternatively one can also consider heavy-tailed matrix ensembles with 
rotational invariance, at the expense of losing statistical independence of 
matrix elements. Both directions have been actively pursued 
\cite{PhysRevE.53.2200,Muttalib:1993zz,
Garcia-Garcia2001,PhysRevE.69.066131,
PhysRevE.70.065102,Burda:2006nk,PhysRevE.76.051105,PhysRevE.77.011122,
Akemann:2008zz,Choi2009,AbulMagd:1900zz}, revealing a plethora of 
exotic behaviors not seen in Gaussian RMTs.  

So far, applications of chRMT to QCD have been mostly limited to the hadronic phase 
with $\akakko{\bar\psi\psi}\ne 0$. Extensions to the high-density regime where the diquark condensate 
$\akakko{\psi\psi}\ne 0$ preponderates were explored in 
\cite{Kanazawa:2009en,Akemann:2010tv,Kanazawa:2011tt,Kanazawa:2014lga}. 
While spontaneous symmetry breaking in all these cases can be 
characterized by a nonvanishing fermion \emph{bilinear} condensate, 
symmetry breaking in general can also be triggered by higher-order condensates. 
In QCD it was stressed by Stern that 
chiral symmetry breaking with $F_\pi\ne0$ does not 
necessitate $\akakko{\bar\psi\psi}\ne 0$  
\cite{Stern:1997ri,Stern:1998dy}. 
Indeed one can imagine a situation where chiral condensate is 
forbidden by an anomaly-free discrete subgroup of $\U(1)_A$ 
and the spontaneous breaking $\SU(N_f)_R\times\SU(N_f)_L\to\SU(N_f)_V$ 
is driven by a \emph{quartic} condensate. (This pattern of symmetry breaking was 
studied by Dashen long time ago \cite{Dashen:1969eg}.) While this exotic phase 
that we call the \emph{Stern phase} is ruled out by rigorous QCD inequalities at vanishing baryon 
density \cite{Kogan:1998zc}, there are arguments in favor of the Stern phase 
at finite density. First, color superconducting phases in dense QCD are examples 
of the Stern phase due to the fact that the leading \emph{gauge-invariant} 
order parameter that breaks chiral symmetry is provided by four-quark condensates 
\cite{Rajagopal:2000wf,Alford:2007xm}. 
Secondly, in phases with spatially modulated chiral condensates, the phonon 
fluctuations associated with translational symmetry breaking wipe out the 
spatial order and lead to a phase with quartic condensates 
\cite{Hidaka:2015xza}. Other related arguments can be found in 
\cite{DescotesGenon:1999zj,Girlanda:2001pc,Watanabe:2003xt,
Harada:2009nq,Adhikari:2011zf,Kanazawa:2015kca}. 

In this paper, we introduce a new heavy-tailed chRMT that corresponds to the Stern phase. 
To be precise, we show that our chRMT with $N\times N$ random matrices reproduces, 
in the large-$N$ limit, the finite-volume partition function of the Stern phase with $K>4$ in the $\epsilon$-regime. 
(Here we label the Stern phase with an index $K$ that specifies the unbroken subgroup of $\U(1)_A$ 
\cite{Kanazawa:2015kca}.) This implies that all infinitely many sum rules for the Dirac eigenvalues in the Stern phase 
\cite{DescotesGenon:1999zj,Kanazawa:2015kca} are obeyed by microscopic eigenvalues of random matrices in the new chRMT.  
In the chiral limit, our chRMT coincides with the model previously considered by Akemann and Vivo \cite{Akemann:2008zz}. 
Here we solve the model at large $N$ with arbitrary quark masses and analytically obtain 
the microscopic spectral density and the smallest eigenvalue distribution in dependence of quark masses. 
We will not compute the macroscopic large-$N$ limit or spectral densities at finite $N$ because they are not universal 
quantities. 

This paper is structured as follows. In Section \ref{sc:RMT} 
we define the model and discuss its relevance to QCD. Then 
we solve the model analytically in the large-$N$ limit and 
obtain the microscopic spectral density and 
the smallest eigenvalue distribution, 
for an arbitrary number of flavors and arbitrary quark masses. 
Section \ref{sc:conc} is devoted to conclusions and outlook.

\section{\label{sc:RMT}Random matrix theory for the Stern phase of QCD}
\subsection[Definition of the model and the large-$N$ limit]
{\boldmath Definition of the model and the large-$N$ limit}

The matrix model considered in this paper is defined 
by the partition function
\ba
	\label{eq:rmt_stern}
	Z^{\rm S}_{N_f}(\{\m_f\}) \equiv 
	\int\limits_{\mathbb{C}^{N\times N}} \hspace{-5pt}  \dd X 
	\frac{1}{(1+\tr X^\dagger X)^{N^2+NN_f+1}}
	\prod_{f=1}^{N_f}\det 
	\bep
		\m_f^*\1_{N} & X \\ -X^\dagger & \m_f \1_{N}
	\eep \,,
\ea
where $X$ is a complex $N\times N$ random matrix and $\dd X$ denotes 
the flat Cartesian measure. This integral converges for 
arbitrary $N\geq 1$ and $N_f\geq 0$. The weight \eqref{eq:rmt_stern} 
evidently has three interesting properties: (i) it is invariant under 
unitary rotations $X\to V_1XV_2$ for $V_{1,2}\in\U(N)$, (ii) the 
matrix elements are statistically correlated, and (iii) the distribution 
is heavy tailed, i.e., it does not decay exponentially for large matrix 
elements. 
This random matrix ensemble 
can be seen as an unquenched generalization of 
previous RMTs \cite{PhysRevE.69.066131,PhysRevE.70.065102,
Burda:2006nk,PhysRevE.77.011122,Akemann:2008zz,AbulMagd:1900zz} 
that had a heavy-tailed weight similar to 
\eqref{eq:rmt_stern} but with no determinants. 
If the weight for $X$ in \eqref{eq:rmt_stern} is replaced with a Gaussian distribution, 
the model reverts to the standard $\beta=2$ chRMT 
called the chiral Gaussian unitary ensemble (chGUE) 
\cite{Shuryak:1992pi,Verbaarschot:1993pm}.  

In the large-$N$ limit, our chRMT enjoys a sigma-model 
representation. To see this, we rewrite \eqref{eq:rmt_stern} 
up to a trivial multiplicative constant as
\ba
	Z^{\rm S}_{N_f}(\{\m_f\}) 
	& \sim 
	\int\limits_{\mathbb{C}^{N\times N}} \hspace{-5pt}  \dd X 
	\int_\mathbb{C} \dd^2z\,\ee^{-N(1+\tr X^\dagger X)|z|^2}|z|^{2N^2+2NN_f}
	\prod_{f=1}^{N_f}\det 
	\bep
		\m_f^*\1_N & X \\ - X^\dagger & \m_f \1_N
	\eep
	\notag 
	\\
	& = 
	\int_\mathbb{C} \dd^2z\,\ee^{-N|z|^2} \hspace{-8pt} 
	\int\limits_{\mathbb{C}^{N\times N}} \hspace{-5pt} 
	\dd X\ |z|^{2N^2} \ee^{-N|z|^2\tr X^\dagger X}
	\prod_{f=1}^{N_f}\det 
	\bep
		z^*\m_f^* \1_N & zX \\ - z^*X^\dagger & z \m_f \1_N
	\eep 
	\notag 
	\\
	& = \int_\mathbb{C} \dd^2z\,\ee^{-N|z|^2} \hspace{-8pt} 
	\int\limits_{\mathbb{C}^{N\times N}} \hspace{-5pt} \dd W
	\ee^{-N\tr W^\dagger W} \prod_{f=1}^{N_f}\det 
	\bep
		z^* \m_f^*\1_N & W \\ -W^\dagger & z \m_f \1_N
	\eep\,,
	\label{eq:Zv2}
\ea
where in the last step we introduced $W\equiv zX$. 
The final expression \eqref{eq:Zv2} is akin to 
the normal chGUE except that the mass term is multiplied by 
another Gaussian random variable $z$ of order $1/\sqrt{N}$. 
This implies that we need a large-$N$ limit with $\m_f=\calO(1/\sqrt{N})$, 
which is different from the conventional large-$N$ limit 
with $\m_f=\calO(1/N)$. Following the standard 
route of bosonization \cite{Shuryak:1992pi}, one can easily obtain
\ba
	Z^{\rm S}_{N_f}(\{\m_f\}) & \sim 
	\int_\mathbb{C} \dd^2z\,\ee^{-N|z|^2} \hspace{-7pt} 
	\int\limits_{\U(N_f)}\!\!\! \dd U\ 
	\exp\kkakko{N\tr ( z\M U + z^*\M^\dagger U^\dagger) }
	\\
	& = \int\limits_{\SU(N_f)}\!\!\! \dd U \  
	\exp\kkakko{N\tr (\M U) \tr (\M^\dagger U^\dagger)}\,,
	\label{eq:EFT_stern}
\ea
where $\dd U$ denotes the Haar measure and $\M\equiv 
\diag(\m^{}_1,\dots,\m^{}_{N_f})$. 
Equation \eqref{eq:EFT_stern} exactly 
coincides with the $\epsilon$-regime finite-volume partition function of QCD 
in the Stern phase with $K>4$ \cite{Kogan:1998zc,DescotesGenon:1999zj,
Kanazawa:2015kca}.  Notably, 
the sigma model \eqref{eq:EFT_stern} has no term at $\calO(\M)$ 
in the exponent, in contradistinction to the standard chiral 
Lagrangian.%
\footnote{Attempts to recover higher-order terms of 
chiral perturbation theory from chRMT have been made 
in \cite{Damgaard:2010cz,Osborn:2010eq} with the 
purpose of studying lattice fermions, which is totally different from our motivation.} 
This reflects that the chiral condensate 
$\akakko{\bar\psi\psi}\to 0$ in the chiral limit. 
We find it intriguing that the sigma model structure has changed in the absence of  
any changes in symmetries of the underlying random matrix. 
The new chRMT may serve as a toy model for 
spontaneous symmetry breaking driven by higher-order condensates.  
It follows from the coincidence of mass dependence between chRMT and QCD 
that infinitely many spectral sum rules of the Dirac operator 
in the Stern phase \cite{DescotesGenon:1999zj,Kanazawa:2015kca} 
can be reproduced exactly from chRMT.
This suggests that the universal behavior of small Dirac eigenvalues 
originating from chiral symmetry breaking could be probed by using this chRMT, which 
shares the same pattern of symmetry breaking as the Stern phase of QCD 
but is much simpler and analytically tractable. 
We end this subsection with two supplementary remarks. 
\begin{itemize}
	\item 
	The partition function \eqref{eq:EFT_stern} has no dependence on 
	the gauge-field topology.  In the Stern phase with $K>4$, topological sectors 
	with nonzero winding numbers are suppressed in the leading order of the 
	$\epsilon$ expansion \cite{Kanazawa:2015kca}. This deprives us of a 
	physical motivation to study the model \eqref{eq:rmt_stern} 
	with rectangular $X$. Nevertheless, it could be mathematically interesting 
	to investigate such extensions in future work.%
	\footnote{For $N_f=0$, this task has already been undertaken 
	in \cite{Akemann:2008zz}. } 
	\item
	The sign of the leading term in the exponent 
	of \eqref{eq:EFT_stern} was fixed unambiguously by chRMT, despite that 
	both signs are allowed by symmetries. 
	Actually, the sign of the leading term \emph{can} be flipped 
	if we modify the fermion determinant in \eqref{eq:Zv2} as 
	\ba
	    \det
	    \bep
	      z^* \m_f^*\1_N & W \\ -W^\dagger & z \m_f \1_N
	    \eep
	    \to ~\det
	    \bep
	      z^* \m_f^* \1_N & W \\ W^\dagger & z \m_f \1_N
	    \eep
	    = |z|^{2N} \det
	    \bep
	      \m_f^* \1_N & W/z \\ W^\dagger/z^* & \m_f \1_N
	    \eep\,.
	\ea
	However the Dirac operator is now \emph{Hermitian}! 
	This means that the anti-Hermiticity of the Dirac operator imposes 
	a nontrivial constraint on the sign of the low energy constant. 
	A similar observation was made in chRMT for Wilson fermions 
	\cite{Akemann:2010em}.
\end{itemize}

\subsection{Microscopic spectral density}
\label{sc:rho_micro}

When $^\forall \m_f\in\RR$, the partition function \eqref{eq:Zv2} becomes 
\begin{gather}
	\label{eq:Zssse}
	Z^{\rm S}_{N_f}(\M) = 
	\int_\mathbb{C} \dd^2z\,\ee^{-N|z|^2} 
	|z|^{2NN_f} \hspace{-5pt} 
	\int\limits_{\mathbb{C}^{N\times N}} \hspace{-5pt} \dd W 
	\ee^{-N\tr W^\dagger W}
	\prod_{f=1}^{N_f}\det (\m_f \1_{2N} + \D_{\rm S})\,,
	\intertext{with the Dirac operator}
	\D_{\rm S} \equiv \frac{1}{|z|}\bep {\bf 0} & W \\ -W^\dagger & {\bf 0} \eep .
\end{gather}
Our primary interest is in the spectral statistics 
of $\D_{\rm S}$ on the scale $1/\sqrt{N}$. 
In this ``microscopic domain'', the eigenvalue density and eigenvalue correlations 
are expected to be \emph{universal} in the sense that it is solely determined 
by the pattern of global symmetry breaking, with no dependence on specific details 
of UV interactions in QCD that cause the symmetry breaking. 
In the following, we derive the microscopic spectral density in the large-$N$ limit, 
first in the chiral limit (Sec.~\ref{sc:rho_chiral_lim}) 
and then for arbitrary quark masses (Sec.~\ref{sc:massive}), by making 
use of a formal similarity of \eqref{eq:Zssse} to the chGUE.  
Our notation is fixed as follows. 
\begin{itemize}
	\item 
	$\ckakko{\pm i\lambda^{\rm S}_n}_{n=1}^{N}$ 
	(with $^\forall\lambda^{\rm S}_n\geq 0$) 
	$~~{\bf\cdots}~~$  Eigenvalues of $\D_{\rm S}$ 
	\item 
	$\ckakko{\pm i\lambda_n}_{n=1}^{N}$ 
	(with $^\forall \lambda_n\geq 0$) 
	$~~{\bf\cdots}~~$ Eigenvalues of 
	$\displaystyle \bep {\bf 0} & W \\ -W^\dagger & {\bf 0} \eep$ 
\end{itemize} 
Obviously, $\displaystyle \lambda_n^{\rm S}=\frac{\lambda_n}{|z|}$ 
for every $n$.

\subsubsection{Chiral limit}
\label{sc:rho_chiral_lim}

Let us begin with the chiral limit. The spectral density of $\D_{\rm S}$ 
at finite $N$ is defined as
\ba
	  R^{\rm S}_{N, N_f}(\lambda) & \equiv 
	  \left\langle \sum\limits_{n=1}^{N} 
	  \delta(\lambda-\lambda^{\rm S}_n) \right\rangle
	  \\
	  & = 
	  \frac{\displaystyle \int_\mathbb{C} \dd^2z\,\ee^{-N|z|^2} 
	  |z|^{2NN_f}  \int \dd W \ee^{-N\tr W^\dagger W}
	  \left[\sum\limits_{n=1}^{N} \delta(\lambda-\lambda^{\rm S}_n) \right]
	  \prod_{f=1}^{N_f}\det \D_{\rm S}
	  }{ \displaystyle 
	  \int_\mathbb{C} \dd^2z\,\ee^{-N|z|^2} 
	  |z|^{2NN_f} \int \dd W
	  \ee^{-N\tr W^\dagger W}  \prod_{f=1}^{N_f}\det \D_{\rm S}
	  }
	  \\
	  & = 
	  \frac{\displaystyle 
	  \int_\mathbb{C} \dd^2z\,\ee^{-N|z|^2} 
	  \int \dd W  \ee^{-N\tr W^\dagger W}
	  \left[\sum\limits_{n=1}^{N} 
	  \delta \mkakko{\lambda-\frac{\lambda_n}{|z|}} \right]
	  {\det}^{N_f} W^\dagger W
	  }{ \displaystyle 
	  \int_\mathbb{C} \dd^2z\,\ee^{-N|z|^2}  \int \dd W 
	  \ee^{-N\tr W^\dagger W}  {\det}^{N_f} W^\dagger W
	  }
	  \label{eq:R_S_chiral}
	  \\
	  & = 
	  \frac{\displaystyle \int_\mathbb{C} 
	  \dd^2z\,\ee^{-N|z|^2} |z|\, R_{N,N_f}(|z|\lambda)}
	  {\displaystyle \int_\mathbb{C} \dd^2z\,\ee^{-N|z|^2}}\,,
\ea
where we have introduced the spectral density in massless chGUE
\ba
	R_{N,N_f}(\lambda) \equiv 
	\frac{\displaystyle 
		\int \dd W \ee^{-N\tr W^\dagger W}
		\kkakko{
			\sum_{n=1}^{N}\delta(\lambda - \lambda_n) 
		}
		{\det}^{N_f} W^\dagger W 
	}{\displaystyle 
		\int \dd W \ee^{-N\tr W^\dagger W}
		{\det}^{N_f} W^\dagger W
	}\,. 
	\label{eq:Rnn}
\ea
The microscopic limit of \eqref{eq:Rnn} was derived in \cite{Verbaarschot:1993pm}. 
Now we use this result to obtain the microscopic spectral density for the new chRMT,
\ba
	  \rho^{\rm S}_{N_f}(\zeta) & \equiv \lim_{N\to\infty}
	  \frac{1}{\sqrt{N}}R^{\rm S}_{N,N_f}\left(\frac{\zeta}{\sqrt{N}}\right)
	  \label{eq:rhoSm0}
	  \\
	  & = \int_\mathbb{C} \frac{\dd^2z}{\pi}\,\ee^{-|z|^2} |z| 
	  \lim_{N\to\infty} \frac{1}{N} R_{N,N_f}\!\!\left(\frac{|z|\zeta}{N}\right)
	  \\
	  & = 4 \int_0^\infty \!\!\dd x~x^2 \ee^{-x^2} 
	  \rho^{\mathstrut}_{N_f}(2x\zeta) \,, 
	  \label{eq:rhosss}
\ea
where we have introduced the microscopic spectral density 
for massless chGUE \cite{Verbaarschot:1993pm}
\ba
	\label{eq:rhoold}
	\rho^{\mathstrut}_{N_f}(\zeta) \equiv
	\frac{\zeta}{2} \Big( J_{N_f}^2(\zeta)-J_{N_f+1}(\zeta)J_{N_f-1}(\zeta) \Big).  
\ea
The integral in \eqref{eq:rhosss} can be performed analytically, 
resulting in a compact expression
\ba
	  \rho^{\rm S}_{N_f}(\zeta) = 2\zeta \ee^{-2\zeta^2}
	  I^{\mathstrut}_{N_f}(2\zeta^2)\,. 
	  \label{eq:rho_massless}
\ea
Here $I_n(z)$ is the modified Bessel function of the first kind. 
\begin{figure}[tb]
	  \begin{center}
	  	\hspace{40pt}
	  	\includegraphics[width=9cm]{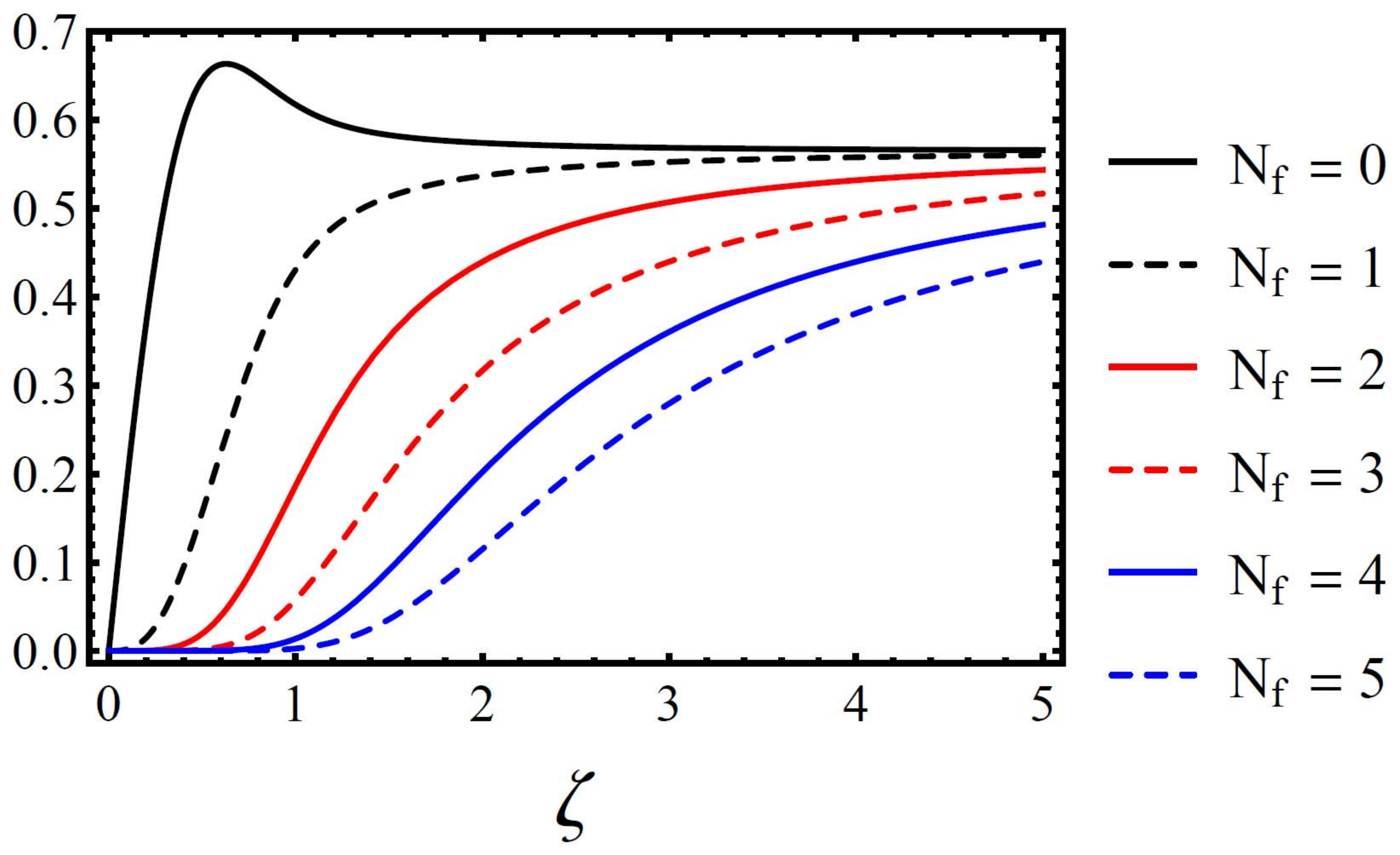}
	  	\put(-298, 91){\large $\rho_{N_f}^{\rm S}(\zeta)$}
	  \end{center}
	  \vspace{-1.5\baselineskip}
	  \caption{\label{fg:stern_exact} 
	  	Microscopic spectral density \eqref{eq:rho_massless} 
	  	in the chiral limit for various $N_f$.
	  }
\end{figure}%
In Fig.~\ref{fg:stern_exact} we show $ \rho^{\rm S}_{N_f}(\zeta)$ 
for several values of $N_f$. It converges to 
$1/\sqrt{\pi}=0.564\dots$ as $\zeta\to\infty$.  As $N_f$ increases, the density 
of eigenvalues near zero is depleted because of the determinant 
in the measure. Comparing $\rho^{\rm S}_{N_f}(\zeta)$ 
with $\rho^{\mathstrut}_{N_f}(\zeta)$, 
we notice that $\rho^{\rm S}_{N_f}(\zeta)$ is flat and monotonic (except for 
$N_f=0$), showing no oscillatory behavior typical of $\rho_{N_f}(\zeta)$. Let us recall that, 
in the chGUE, the oscillation is produced by peaks in the density of individual small eigenvalues.  
By contrast, as we will see in Sec.~\ref{sc:Es}, the density of individual eigenvalues near zero 
in our chRMT is so broad that their superposition smears out each peak completely. 
Aside from this difference, $\rho^{\mathstrut}_{N_f}(\zeta)$ and 
$\rho^{\rm S}_{N_f}(\zeta)$ 
look similar, but once again we emphasize that 
$\rho^{\rm S}_{N_f}(\zeta)$ is the density at the scale 
$\lambda_n^{\rm S}\sim 1/\sqrt{N}$, whereas $\rho^{\mathstrut}_{N_f}(\zeta)$ is 
the density at the scale $\lambda_n \sim 1/N$ --- these two regimes are 
totally different. It was originally pointed out by Stern \cite{Stern:1997ri,Stern:1998dy} 
that chiral symmetry could be spontaneously broken when near-zero Dirac 
eigenvalues scale as $1/\sqrt{V_4}$ instead of $1/V_4$ 
in the thermodynamic limit $V_4\to \infty$. Our finding within chRMT is  
fully consistent with Stern's perspective. While the Stern phase is ruled out  
by rigorous QCD inequalities in QCD at zero density \cite{Kogan:1998zc}, its realization 
in zero-dimensional chRMT is not prohibited. As a side remark, we mention that 
a universal behavior of Dirac eigenvalues on the scale $1/\sqrt{V_4}$ is known 
in strongly non-Hermitian chRMTs corresponding to the BCS regime of dense 
QCD-like theories \cite{Yamamoto:2009ey,Kanazawa:2009en,Akemann:2010tv,Kanazawa:2012zzr,Kanazawa:2014lga}. 

Usually one associates the approach of $\rho^{\mathstrut}_{N_f}(\zeta)$ 
to a constant value at $\zeta\to\infty$ with a nonzero chiral condensate 
through the Banks-Casher relation \cite{Banks:1979yr}. 
It must be noted, however, that the same behavior of 
$\rho^{\rm S}_{N_f}(\zeta)$ does \emph{not} imply a nonvanishing chiral 
condensate. The reason is that, in this model,  
the height of the macroscopic spectral density $R^{\rm S}_{N,N_f}(\lambda)$ 
at the origin scales as $\sqrt{N}$ for $N\gg 1$, 
implying that the chiral condensate $\displaystyle\lim_{\lambda\to 0}\lim_{N\to\infty}
\frac{1}{N}R^{\rm S}_{N,N_f}(\lambda)$ vanishes as 
$\displaystyle\propto\frac{1}{\sqrt{N}}$\,. 

An important remark on a preceding work is in order. 
In \cite{Akemann:2008zz}, with a mathematical motivation, 
Akemann and Vivo studied a deformed 
Wishart-Laguerre ensemble, which is essentially equivalent to 
\eqref{eq:rmt_stern} with $N_f=0$. They derived the microscopic 
spectral density in the large-$N$ limit analytically, not only for square $X$ but also for 
rectangular $X$ of size $(N+\nu)\times N$. In addition, their analysis exhausted 
all the three symmetry classes with Dyson index $\beta=1,2$ and $4$.   
Intriguingly, one can show that their model with $\nu>0$ and $\beta=2$ 
exactly coincides with our unquenched model \eqref{eq:rmt_stern} 
with $N_f=\nu$ and $^\forall \m_f=0$. This is a manifestation of what is known in chRMT 
as the duality between flavor and topology \cite{Verbaarschot:1997bf}.  
As a result, \eqref{eq:rhosss} above can be obtained from 
\cite[eq.(4.7)]{Akemann:2008zz} by letting $\nu=N_f$ and sending $\alpha\to 0$ 
there. We confirmed that our results agree with \cite{Akemann:2008zz}. 
Nonetheless we have presented the full derivation above, firstly because the compact 
expression \eqref{eq:rho_massless} is new, and secondly because 
the computation in the chiral limit is a useful preliminary step for the generalization 
to the case of arbitrary nonzero quark masses, which is a genuinely new result of this paper 
and will be worked out in the next subsection.

\subsubsection{Nonzero masses}
\label{sc:massive}

\begin{figure}[tb]
	  \begin{center}
	  	\hspace{80pt}
	  	\includegraphics[height=55mm]{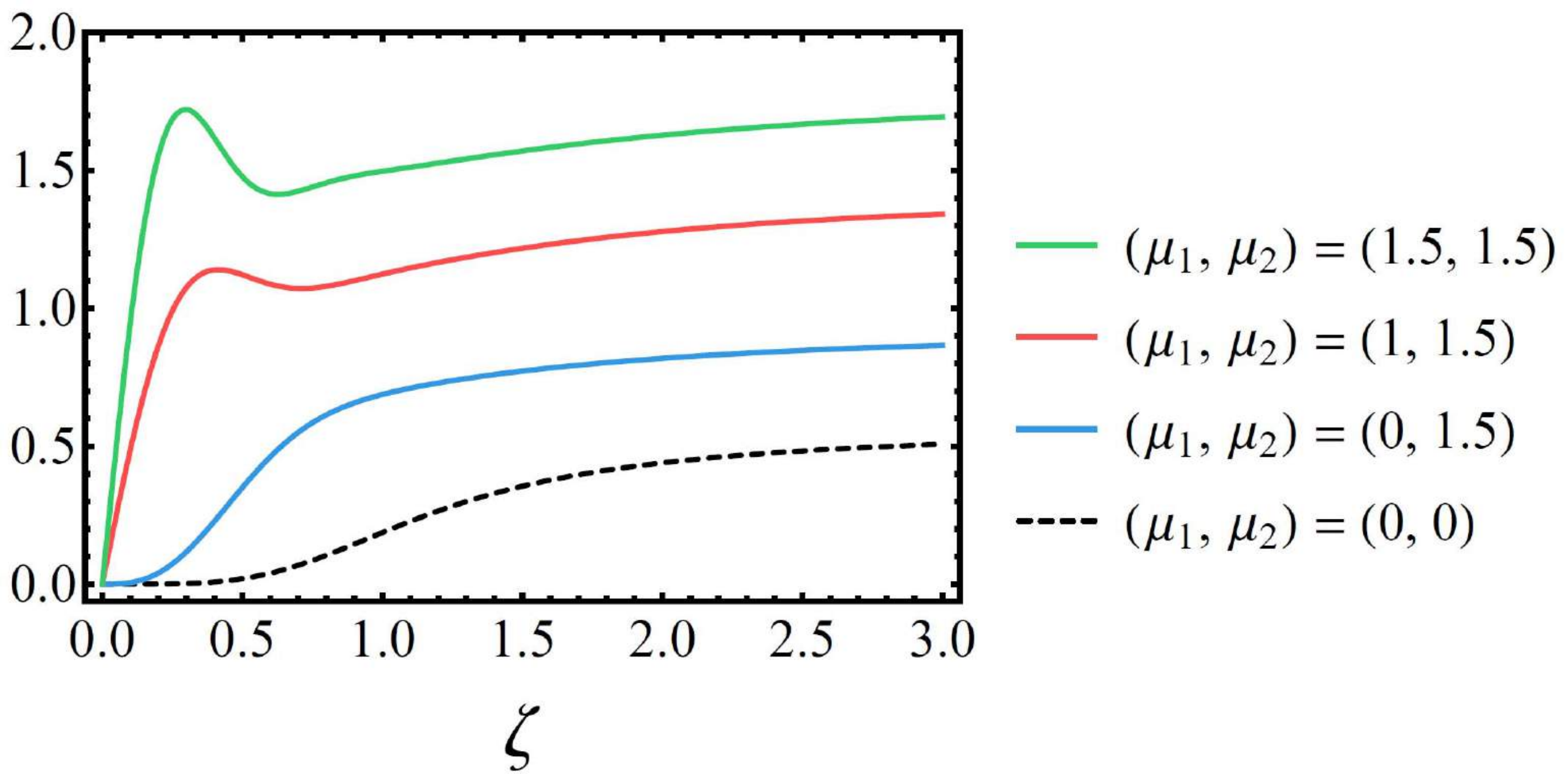}
	  	\put(-394, 92){\large $\rho_{2}^{\rm S}
	  	(\zeta,\{\mu_1,\mu_2\})$}
	  \end{center}
	  \vspace{-1.5\baselineskip}
	  \caption{\label{fg:rho_Nf=2_massive} 
	  	Microscopic spectral density for $N_f=2$. 
	  }
\end{figure}

Reinstating quark masses in \eqref{eq:R_S_chiral} 
and replacing $z$ by $z/\sqrt{N}$, we obtain
\ba
	  & \qquad R^{\rm S}_{N, N_f}(\lambda,\{\m_f\}) 
	  \notag
	  \\
	  & = 
	  \frac{\displaystyle 
	  \int_\mathbb{C} \dd^2z\,\ee^{-|z|^2} \int \dd W 
	  \ee^{-N\tr W^\dagger W}
	  \left[
	  	\sum_{n=1}^{N} 
	  	\delta\mkakko{\lambda-\frac{\sqrt{N}\lambda_n}{|z|}} 
	  \right]
	  \prod_{f=1}^{N_f}
	  \det \bigg( \frac{|z|^2\m_f^2}{N}\1_N + W^\dagger W \bigg)
	  }{ \displaystyle 
	  \int_\mathbb{C} \dd^2z\,\ee^{-|z|^2} \int \dd W 
	  \ee^{-N\tr W^\dagger W}  \prod_{f=1}^{N_f}
	  \det \bigg( \frac{|z|^2\m_f^2}{N}\1_N + W^\dagger W \bigg)
	  }
	  \\
	  & = \frac{1}{\sqrt{N}}
	  \frac{\displaystyle 
	  \int_0^\infty \!\! \dd x\,x^2 \ee^{-x^2} \!\!\! \int \dd W 
	  \ee^{-N\tr W^\dagger W}
	  \left[
	  	\sum_{n=1}^{N} \delta\mkakko{\frac{x\lambda}{\sqrt{N}} - \lambda_n}
	  \right]
	  \prod_{f=1}^{N_f}\det\bigg( \frac{x^2\m_f^2}{N}\1_N + W^\dagger W \bigg)
	  }{ \displaystyle 
	  \int_0^\infty \!\!\dd x\,x\ee^{-x^2} \!\!\! \int \dd W 
	  \ee^{-N\tr W^\dagger W}  \prod_{f=1}^{N_f}
	  \det \bigg( \frac{x^2\m_f^2}{N}\1_N + W^\dagger W \bigg)
	  }\,.
\ea
The microscopic limit is achieved by sending $N$ to infinity with 
$\sqrt{N}\lambda_n^{\rm S}\sim\sqrt{N}\m_f\sim \calO(1)$. 
Extending \eqref{eq:rhoSm0} to nonzero masses, we obtain
\ba
	  & \qquad \rho^{\rm S}_{N_f}(\zeta,\{\mu_f\}) 
	  \notag
	  \\
	  & \equiv \lim_{N\to\infty}\frac{1}{\sqrt{N}}R^{\rm S}_{N,N_f}
	  \left(   \frac{\zeta}{\sqrt{N}}, \left\{\frac{\mu_f}{\sqrt{N}}\right\}   \right)
	  \\
	  & = \lim_{N\to\infty} \frac{1}{N}  \frac{\displaystyle 
	  \int_0^\infty \!\! \dd x\,x^2 \ee^{-x^2}  
	  \!\!\! \int \dd W \ee^{-N\tr W^\dagger W} 
	  \left[
	  	  \sum_{n=1}^{N} \delta\mkakko{\frac{x\zeta}{N} - \lambda_n} 
	  \right]
	  \prod_{f=1}^{N_f}\det
	  \kkakko{ \mkakko{\frac{x\mu_f}{N}}^2 \1_N + W^\dagger W }
	  }{ \displaystyle 
	  \int_0^\infty \!\!\dd x\,x\ee^{-x^2} \!\!\! 
	  \int \dd W \ee^{-N \tr W^\dagger W} \prod_{f=1}^{N_f}
	  \det \kkakko{ \mkakko{\frac{x\mu_f}{N}}^2\1_N + W^\dagger W }
	  }
	  \\
	  & = \lim_{N\to\infty}  \frac{\displaystyle 
	    \int_0^\infty \!\! \dd x\,x^2\ee^{-x^2} S_{N,N_f}\left(\left\{\frac{x\mu_f}{N}\right\}\right) 
	    \frac{1}{N}R_{N,N_f}\left(\frac{x\zeta}{N},\left\{\frac{x\mu_f}{N}\right\}\right)
	  }{\displaystyle 
	    \int_0^\infty \!\! \dd x\,x\ee^{-x^2} S_{N,N_f}\left(\left\{\frac{x\mu_f}{N}\right\}\right)
	  }  \,, 
	  \label{eq:rhomassi}
\ea
where we have introduced the partition function for chGUE 
(cf.~\cite{Brower:1981vt,Jackson:1996jb,Balantekin:2000vn})
\ba
	  \label{eq:S_eig_rep}
	  S_{N,N_f}\left(\left\{\frac{x}{N}\mu_f\right\}\right) & 
	  \equiv \int \dd W \ee^{-N\tr W^\dagger W} \prod_{f=1}^{N_f}
	  \det\kkakko{ \mkakko{\frac{x\mu_f}{N}}^2 \1_N + W^\dagger W }
	  \\
	  & \sim \int_{\U(N_f)}
	  \dd U\,\exp\big[ 2x\ \Re\tr(\bm{\mu} U) \big]
	  \qquad \text{for}~~N\gg 1
	  \\
	  &  \! \propto 
	  \frac{1}{\Delta_{N_f}(\{-(2x\mu_f)^2\})} \underset{1\leq i,\,j\leq N_f}{\det}
	  \big[  (2x\mu_j)^{i -1}I_{i-1}(-2x\mu_j)  \big] 
	  \label{eq:bunpai}
\ea
with 
\ba
	\bm{\mu}\equiv \diag(\mu_1,\dots,\mu_{N_f})
	~~~\text{and}~~~
	\Delta_{N_f}(\{a_i\})\equiv \prod_{i>j}(a_i-a_j)\,,
\ea 
whereas the spectral density for massive chGUE is given by
\ba
	\label{eq:4regfddsf}
	R_{N,N_f}(\lambda, \{\m_f\}) & \equiv 
	\frac{\displaystyle 
		\int \dd W \ee^{-N\tr W^\dagger W}
		\kkakko{  \sum_{n=1}^{N} \delta( \lambda - \lambda_n )  }
		\prod_{f=1}^{N_f}\det(\m_f^2 \1_N + W^\dagger W)
	}{\displaystyle 
		\int \dd W \ee^{-N\tr W^\dagger W}
		\prod_{f=1}^{N_f}\det(\m_f^2 \1_N + W^\dagger W)
	}\,. 
\ea
The microscopic limit of \eqref{eq:4regfddsf} was computed 
in \cite{Damgaard:1997ye,Wilke:1997gf} as  
\ba
	  \label{eq:R_standard_limit}
	  \lim_{N\to\infty} \frac{1}{N}R_{N,N_f}
	  \left( \frac{x\zeta}{N};\left\{\frac{x\mu_f}{N}\right\} \right) 
	  = 
	  2 \rho^{\mathstrut}_{N_f}(2x\zeta,\{2x\mu_f\})  
\ea
with
\ba
	  \label{eq:rho_old_massive}
	  \rho^{\mathstrut}_{N_f}(z,\{\m_f\}) \equiv -\frac{1}{2}
	  \frac{\displaystyle 
	    \det \begin{bmatrix}  
	    J_{-1}(z) & z J_0(z)& \cdots & z^{N_f+1}J_{N_f}(z) 
	    \\
	    J_0(z) & z J_1(z) & \cdots &  z^{N_f+1}J_{N_f+1}(z)
	    \\
	    I_0(-\m_1) & \m_1 I_1(-\m_1) & \cdots & \m_1^{N_f+1}I_{N_f+1}(-\m_1) 
	    \\
	    \vdots & \vdots & \ddots & \vdots 
	    \\
	    I_0(-\m^{}_{N_f}) & \m^{}_{N_f}I_1(-\m^{}_{N_f}) 
	    & \cdots & \m_{N_f}^{N_f+1}I_{N_f+1}(-\m^{}_{N_f})  
	    \end{bmatrix}
	  }{\displaystyle 
	    \prod_{f=1}^{N_f}(z^2+\m_f^2)
	    \underset{1\leq i,\,j\leq N_f}{\det}
	    \big[  \m_j^{i -1}I_{i-1}(-\m_j)  \big]
	  }\,.
\ea
In the chiral limit \eqref{eq:rho_old_massive} reduces to \eqref{eq:rhoold}. Now, 
substituting \eqref{eq:bunpai} and \eqref{eq:R_standard_limit} into \eqref{eq:rhomassi},  
we finally arrive at the microscopic spectral density of $\D_{\rm S}$ 
with $N_f$ massive flavors, 
\ba
	  \rho^{\rm S}_{N_f}(\zeta,\{\mu_f\}) 
	  & = - \frac{2}{\displaystyle \prod_{f=1}^{N_f}(\zeta^2+\mu_f^2)} 
	  \frac{\displaystyle 
	    \int_0^\infty \!\!\dd x\,\ee^{-x^2} x^{(3-N_f)(2+N_f)/2} 
	    \det \Big[  \Xi_{N_f}(x,\zeta,\{\mu_f\})  \Big] 
	  }{\displaystyle 
	    \int_0^\infty \!\!\dd x\,\ee^{-x^2} x^{-N_f(N_f-1)/2+1} 
	    \underset{1\leq i,\,j\leq N_f}{\det}
	    \big[  \mu_j^{i -1}I_{i-1}(-2x\mu_j)  \big]
	  }\,,
	  \label{eq:rho_massive_final}
\ea
with 
\ba
	  \Xi_{N_f}(x,\zeta,\{\mu_f\}) \equiv 
	  \begin{bmatrix}  J_{-1}(2x\zeta) & \zeta J_0(2x\zeta)& \cdots & \zeta^{N_f+1}J_{N_f}(2x\zeta) 
	    \\
	    J_0(2x\zeta) & \zeta J_1(2x\zeta) & \cdots &  \zeta^{N_f+1}J_{N_f+1}(2x\zeta)
	    \\
	    I_0(-2x\mu_1) & \mu_1 I_1(-2x\mu_1) & \cdots 
	    & \mu_1^{N_f+1}I_{N_f+1}(-2x\mu_1) 
	    \\
	    \vdots & \vdots & \ddots & \vdots 
	    \\
	    I_0(-2x\mu^{}_{N_f}) & \mu^{}_{N_f}I_1(-2x\mu^{}_{N_f}) & 
	    \cdots & \mu_{N_f}^{N_f+1}I_{N_f+1}(-2x\mu^{}_{N_f})  
	    \end{bmatrix}
	    \,.
\ea
Let us examine the simplest case closely. For $N_f=1$, 
\eqref{eq:rho_massive_final} can be simplified to
\ba
	  \rho^{\rm S}_{N_f=1}(\zeta, \mu) 
	  & = 
	  - 4 \frac{\ee^{-\mu^2}}{\zeta^2+\mu^2}
	    \int_0^\infty \!\!\dd x\,\ee^{-x^2} x^{3} \det
	    \begin{bmatrix}  
	      J_{-1}(2x\zeta) & \zeta J_0(2x\zeta)& \zeta^{2}J_{1}(2x\zeta) 
	      \\
	      J_0(2x\zeta) & \zeta J_1(2x\zeta) & \zeta^{2}J_{2}(2x\zeta)
	      \\
	      I_0(-2x\mu) & \mu I_1(-2x\mu)  &  \mu^{2}  I_{2}(-2x\mu) 
	    \end{bmatrix}
	    \,.
	    \label{eq:rho_Nf=1_massive}
\ea
\begin{figure}[tb]
	  \begin{center}
	  	\hspace{40pt}
	  	\includegraphics[height=55mm]{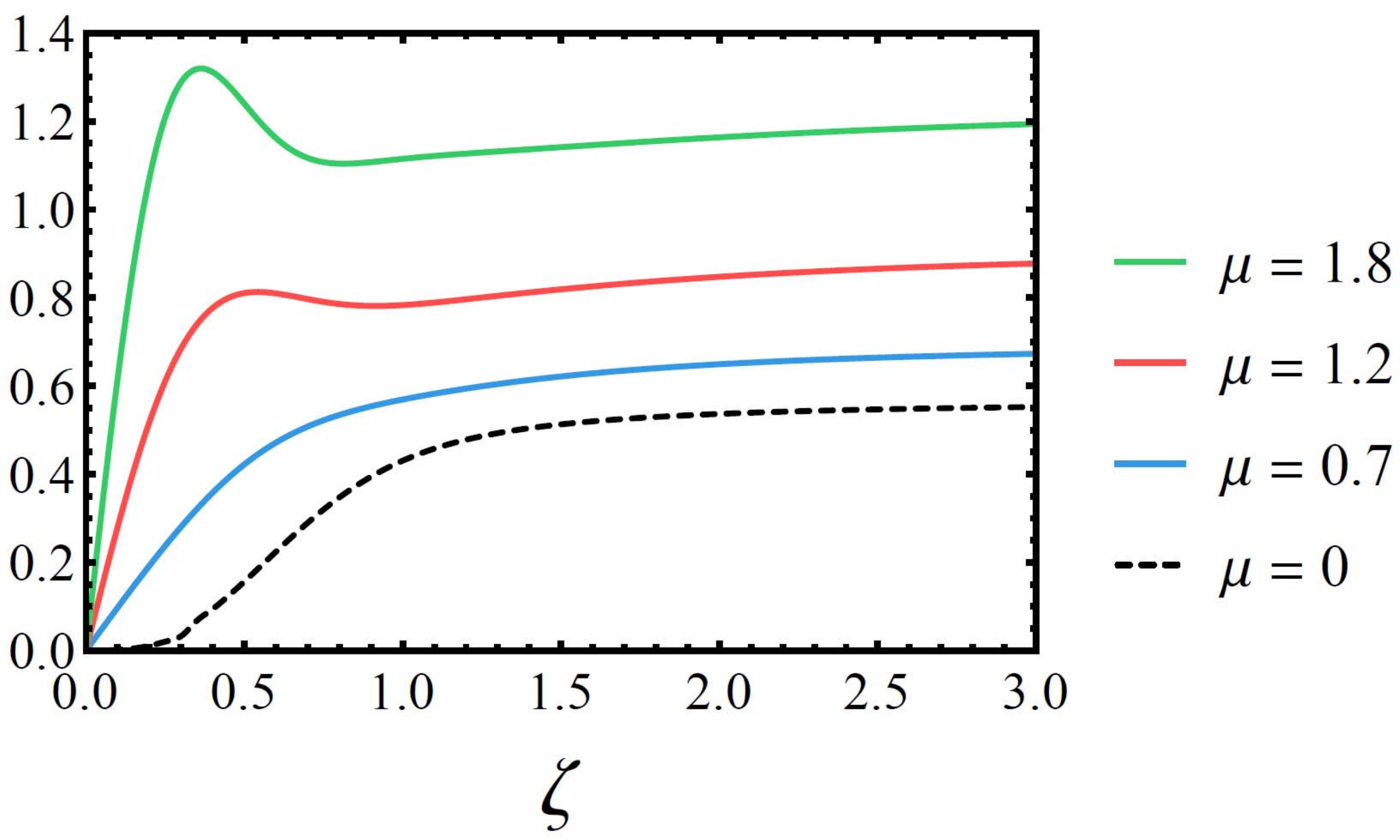}
	  	\put(-307, 90){\large $\rho_{1}^{\rm S}(\zeta,\mu)$}
	  \end{center}
	  \vspace{-1.5\baselineskip}
	  \caption{\label{fg:rho_Nf=1_massive} 
	  	Microscopic spectral density \eqref{eq:rho_Nf=1_massive} 
	  	for $N_f=1$.
	  }
\end{figure}\noindent 
In the limit $\mu\to 0$ \eqref{eq:rho_Nf=1_massive} reproduces  
\eqref{eq:rhosss} for $N_f=1$, as it should. 
We plot $\rho_{N_f=1}^{\rm S}(\zeta,\mu)$ in 
Fig.~\ref{fg:rho_Nf=1_massive} for several values of $\mu$. 
Clearly $\rho_{N_f=1}^{\rm S}(\zeta,\mu)$ increases with $\mu$. 
The asymptotic value at $\zeta\gg 1$ depends on $\mu$, and appears to diverge as 
$\mu\to \infty$. This means that \emph{a heavy flavor does not decouple} --- 
$\rho_{N_f=1}^{\rm S}(\zeta,\mu)$ does not reduce 
to the quenched density at large $\mu$.%
\footnote{Non-decoupling of heavy flavors also occurs in 
non-Hermitian chRMT for dense 
QCD-like theories \cite{Akemann:2010tv,Kanazawa:2012zzr}. 
In this case the origin of non-decoupling is physically understood: 
the Cooper pairing between quarks requires that we send masses of 
an \emph{even number of flavors} to infinity simultaneously. Otherwise 
the Dirac spectrum becomes singular in the infinite-mass limit.}  
This is quite unusual compared to what is known for standard chRMT 
\cite{Damgaard:1997ye,Wilke:1997gf}, where  
the microscopic spectral density approaches $1/\pi$ asymptotically  
for any number of flavors and any masses, and where 
the decoupling of heavy flavors holds in 
the sense that, when some of the masses are sent to infinity, the massive spectral 
density reduces to that for a reduced number of flavors. 
By contrast, Fig.~\ref{fg:rho_Nf=1_massive} reveals that neither 
property persists in the Stern phase. 
For comparison, we also display the massive spectral 
density for $N_f=2$ in Fig.~\ref{fg:rho_Nf=2_massive}, which exhibits 
a similar mass dependence to $N_f=1$. 
We look into this curious behavior in more detail in the next subsection.

\noindent 

\subsubsection{\label{sc:largemass}Large-mass limit}

Why does a heavy flavor fail to decouple? 
Let us examine what happens to the spectral density when 
one of the masses is made large compared to the others. 
Our starting point is the $\epsilon$-regime partition function \eqref{eq:EFT_stern} 
of the Stern phase for $N_f$ light flavors,
\ba
	  Z^{\rm S}_{N_f}(\{\mu_f\}) \sim \int\limits_{\U(N_f)} 
	  \!\!\! \dd U \,\exp\mkakko{|\tr(\bm{\mu}U)|^2}\,.
	  \label{eq:Z_stern_}
\ea
If $\mu^{}_{N_f}$ is by far the largest among $\mu^{}_f$'s, the fluctuation 
of $U$ over $\U(N_f)$ would be effectively restricted to $\U(N_f-1)$, hence 
$U\simeq \bep \tilde U & 0 \\ 0 & 1 \eep$ with $\tilde U\in \U(N_f-1)$. 
By plugging this into \eqref{eq:Z_stern_} and introducing 
a reduced mass matrix $\bm{\mu}_r \equiv 
\diag(\mu_1,\dots,\mu_{N_f-1}^{})$, we get
\ba
	  Z^{\rm S}_{N_f}(\{\mu_f\}) & \sim 
	  \int\limits_{\U(N_f-1)} \!\!\!\!\!\! 
	  \dd U \, \exp\mkakko{|\mu^{}_{N_f}+\tr(\bm{\mu}_r\tilde{U})|^2}
	  \\
	  & \sim \exp\big( \mu^{2\mathstrut}_{N_f} \big) \!\! 
	  \int\limits_{\U(N_f-1)} \!\!\!\!\!\! \dd U\, \exp
	  \kkakko{2\mu^{}_{N_f}
	  \Re\tr(\bm{\mu}_r\tilde{U})}\,,
\ea
which is nothing but the partition function of chGUE 
with $N_f-1$ flavors \cite{Shuryak:1992pi}. Hence one cannot recover 
$Z^{\rm S}_{N_f-1}$ from $Z^{\rm S}_{N_f}$ by sending one of 
the masses to infinity; this is how our naive expectation of decoupling 
fails. Instead, one ends up with the conventional 
chiral Lagrangian with an $\calO(M)$ term whose coefficient is set  
by $\mu^{}_{N_f}$. This implies that a large explicit mass $\mu^{}_{N_f}$ 
induces large chiral condensates $\akakko{\bar\psi_f\psi_f}$ for 
the other $N_f-1$ flavors. Generation of such \emph{induced condensates} 
has been discussed in \cite{Girlanda:2001pc} for large-$N_c$ QCD, 
and our analysis based on chRMT is totally consistent with 
\cite{Girlanda:2001pc}. 

\begin{figure}[bt]
	  \begin{center}
		\includegraphics[height=5.3cm]{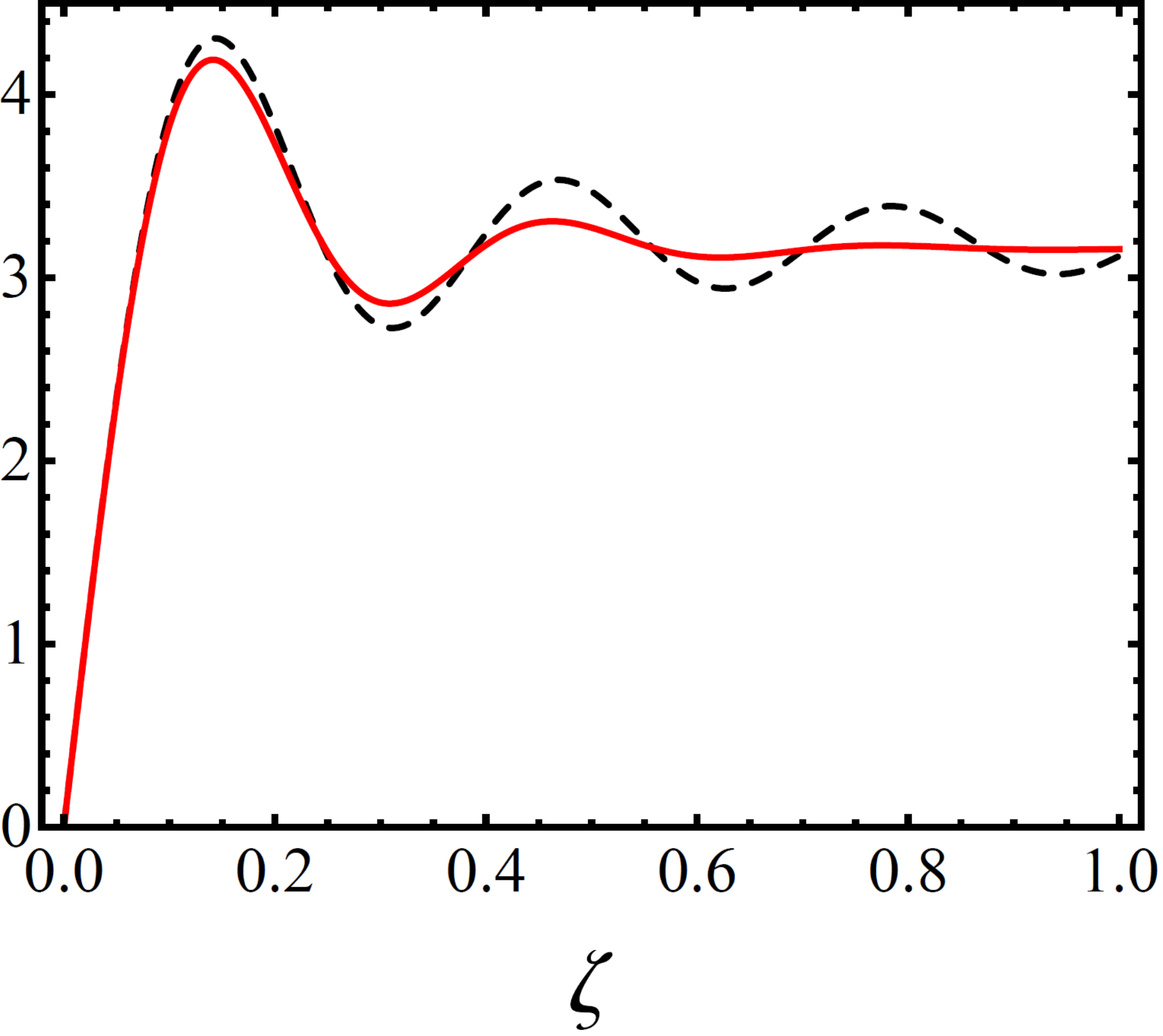}
		\quad \quad 
		\includegraphics[height=5.3cm]{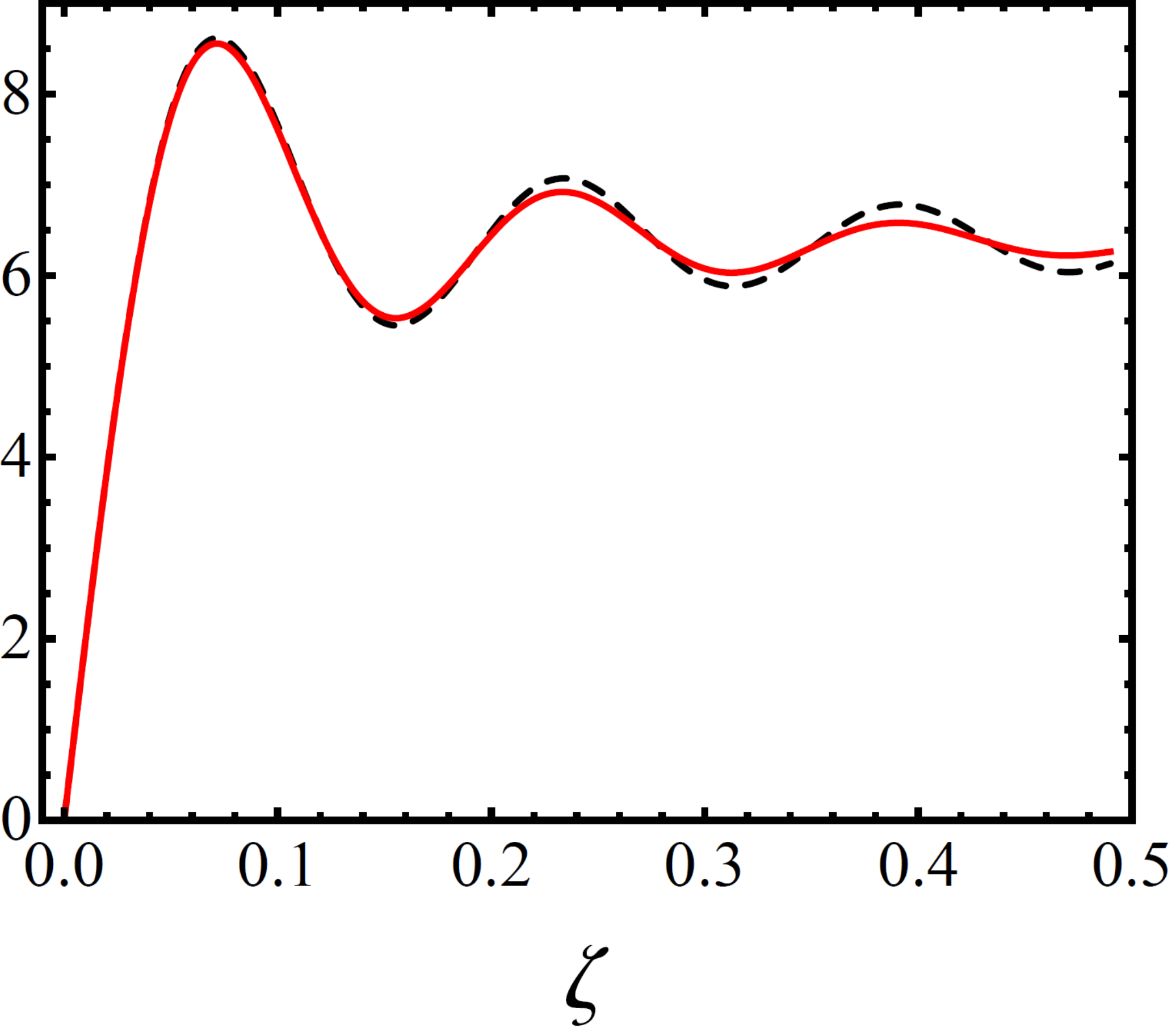}
	  \end{center}
	  \vspace{-1.2\baselineskip}
	  \caption{$\rho^{\rm S}_{N_f=1}(\zeta, \mu)$ 
	  [\eqref{eq:rho_Nf=1_massive}] 
	  for $\mu=5$ (left, red line) and 
	  $\mu=10$ (right, red line) in comparison to 
	  the asymptotic form [RHS of \eqref{eq:rho_asymp_heavy}] 
	  (black dashed lines). 
	  Note the difference of scales in the two figures.
	  }
	  \label{fg:rho_Nf=1_reduction}
\end{figure}

As there is a generic correspondence between sigma models 
and spectral statistics, one can expect that the spectral density 
for the Stern phase at large $\mu^{}_{N_f}$ would reduce to 
that of chGUE whose Gaussian distribution parameter is set by 
$\mu^{}_{N_f}$. In fact, 
when $\mu^{}_{N_f}\gg 1$ and 
$\zeta\sim \mu_f\sim \calO(1/\mu^{}_{N_f}) \ll 1$ 
for $1\leq f\leq N_f-1$, there exists a relation
\ba
	\label{eq:rho_asymp_heavy}
	\rho^{\rm S}_{N_f}(\zeta,\{\mu_f\})
	& \simeq 2\mu^{\mathstrut}_{N_f} 
	\rho^{\mathstrut}_{N_f-1}
	(2\mu^{\mathstrut}_{N_f}\zeta,\{2\mu^{}_{N_f}\mu_f\})\,.
\ea
This can be shown from \eqref{eq:rho_massive_final} by 
using the Laplace expansion of a determinant and approximating 
the modified Bessel function by its asymptotic form. 
The relation \eqref{eq:rho_asymp_heavy} explicitly provides a novel 
link between the spectral density in the Stern phase and that in chGUE.  
To assess the accuracy of \eqref{eq:rho_asymp_heavy}, 
we display $\rho^{\rm S}_{N_f=1}(\zeta,\mu)$ for $\mu=5$ and $10$ 
in Fig.~\ref{fg:rho_Nf=1_reduction}, together with the RHS of 
\eqref{eq:rho_asymp_heavy}. 
While the agreement is good for small $\zeta$, 
deviations emerge for $\zeta \gtrsim 1/\mu$. 
An oscillatory behavior not present in the chiral limit 
gradually sets in as $\mu$ increases.

\noindent 

\subsection{Smallest eigenvalue distribution}
\label{sc:Es}

Next we turn to the smallest eigenvalue distribution in the 
large-$N$ microscopic limit of the chRMT for the Stern phase. 
As in the previous sections, we work with the 
rescaled masses $\mu_f=\sqrt{N}\m_f$. 
We first define the so-called \emph{gap probability} 
\ba
	\label{eq:gagap}
	E^{}_{N,N_f}(\zeta) & \equiv 
	\akakko{
		\prod_{n=1}^{N} 
		\Theta\big(\sqrt{N}\lambda_n^{\rm S} - \zeta \big)
	}\,,
\ea
which is the probability that \emph{none} of 
$\{\sqrt{N}\lambda^{\rm S}_n \}_n$  
falls into the interval $[0, \zeta]$. By definition, 
$\displaystyle\lim_{\zeta \to +0}E^{}_{N,N_f}(\zeta)=1$. 
The factor $\sqrt{N}$ in \eqref{eq:gagap} indicates that 
we are probing the microscopic domain with 
$\lambda_n^{\rm S}\sim N^{-1/2}$. 
The importance of the gap probability stems from the relation
\ba
       E^{}_{N,N_f}(\zeta) = \int_{\zeta}^{\infty}\!\! 
       \dd \zeta_{\min}\ P^{}_{N,N_f}(\zeta_{\min};\{\mu_f\})
\ea
with $P^{}_{N,N_f}$ the smallest eigenvalue distribution. Now, 
by applying the method of \cite{Wilke:1997gf}, it is somewhat 
tedious but straightforward to show 
\ba
	  & P^{}_{N,N_f}(\zeta;\{\mu_f\}) 
	  =  -\frac{\dd}{\dd \zeta}E^{}_{N,N_f}(\zeta)
	  \\
	  \label{eq:Ps_rep_by_R_S}
	  & ~~ = \scalebox{0.9}{$\displaystyle 
	  \lim_{\lambda\to +0} \frac{\zeta}{\lambda} \frac{\displaystyle 
	    \int_0^\infty \!\! \dd x\,x^3 \ee^{-(1+\zeta^2)x^2} 
	    S^{}_{N,N_f}\Big( \Big\{ \frac{x}{N}\sqrt{\mu_f^2+\zeta^2} \Big\} \Big) 
	    \frac{1}{N}R^{}_{N,N_f}\left(\frac{\lambda}{N};
	    \Big\{ \frac{x}{N}\sqrt{\mu_f^2+\zeta^2} \Big\}\right)
	  }{\displaystyle 
	    \int_0^\infty \!\!\dd x\,x \ee^{-x^2} 
	    S^{}_{N,N_f}\left( \left\{ \frac{x}{N}\mu_f \right\} \right)
	  }
	  $}\,,
\ea
where $S^{}_{N,N_f}$ and $R^{}_{N,N_f}$ are the partition function 
and the spectral density of chGUE, respectively, 
as defined in \eqref{eq:S_eig_rep} and \eqref{eq:4regfddsf}. 
Then it is easy to take the large-$N$ microscopic limit by 
exploiting \eqref{eq:bunpai} and \eqref{eq:R_standard_limit}, 
with the result 
\ba
	  & P^{}_{N_f}(\zeta;\{\mu_f\}) 
	  \equiv \lim_{N\to \infty} P^{}_{N,N_f}(\zeta;\{\mu_f\}) 
	  \\
	  &\qquad = 2 \zeta \frac{\displaystyle 
	    \int_0^\infty \!\! \dd x\,x^3 \ee^{-(1+\zeta^2)x^2}
	    \frac{1}{\Delta_{N_f}(\{-(2x)^2(\mu_f^2+\zeta^2)\})}  \ 
	    \Omega^{}_{N_f}\Big(\big\{ 2x\sqrt{\mu_f^2+\zeta^2} \big\}\Big) 
	  }
	  {\displaystyle 
	    \int_0^\infty \!\! \dd x\,x \ee^{-x^2}
	    \frac{1}{\Delta_{N_f}(\{-(2x\mu_f)^2\})} \underset{1\leq i,\,j\leq N_f}{\det}
	    \big[  (2x\mu_j)^{i -1}I_{i-1}(-2x\mu_j)  \big]
	  }
	  \\
	  \label{eq:P_Nf_final}
	  & \qquad = 2\zeta \frac{\displaystyle 
	    \int_0^\infty \!\! \dd x\,x^{3-N_f(N_f-1)} 
	    \ee^{-(1+\zeta^2)x^2}
	    \Omega^{}_{N_f}\Big(\big\{ 2x\sqrt{\mu_f^2+\zeta^2} \big\}\Big) 
	  }
	  {\displaystyle 
	    \int_0^\infty \!\! \dd x\,x^{1-N_f(N_f-1)} \ee^{-x^2}
	    \underset{1\leq i,\,j\leq N_f}{\det}
	    \big[  (2x\mu_j)^{i -1}I_{i-1}(-2x\mu_j)  \big]
	  }\,,
\ea
where 
\ba
	  \Omega^{}_{N_f}(\{\m_f\}) & \equiv 
	  - \lim_{\alpha\to 0}\frac{1}{\alpha} \frac{\displaystyle 
	    \det \begin{bmatrix}  
	    J_{-1}(\alpha) & \alpha J_0(\alpha)& \cdots & \alpha^{N_f+1}J_{N_f}(\alpha) 
	    \\
	    J_0(\alpha) & \alpha J_1(\alpha) & \cdots &  \alpha^{N_f+1}J_{N_f+1}(\alpha)
	    \\
	    I_0(-\m_1) & \m_1 I_1(-\m_1) & \cdots & \m_1^{N_f+1}I_{N_f+1}(-\m_1) 
	    \\
	    \vdots & \vdots & \ddots & \vdots 
	    \\
	    I_0(-\m^{}_{N_f}) & \m^{}_{N_f}I_1(-\m^{}_{N_f}) 
	    & \cdots & \m_{N_f}^{N_f+1}I_{N_f+1}(-\m^{}_{N_f})  
	    \end{bmatrix}
	  }{\displaystyle 
	    \prod_{f=1}^{N_f}(\alpha^2 + \m_f^2)
	  }
	  \\
	  & = \det \begin{bmatrix}
	    I_2(-\m_1) & \cdots & \m_1^{N_f-1}I_{N_f+1}(-\m_1) 
	    \\
	    \vdots & \ddots & \vdots 
	    \\
	    I_2(-\m^{}_{N_f}) 
	    & \cdots & \m_{N_f}^{N_f-1}I_{N_f+1}(-\m^{}_{N_f})  
	  \end{bmatrix}
	  \,.
\ea
This is a new result. 
For small $N_f$, integrals in \eqref{eq:P_Nf_final} 
can be carried out analytically and yield simple expressions: 
\begin{subequations}
	\label{eq:246abc}
	\ba
		P^{}_0(\zeta) & = \frac{2\zeta}{(1+\zeta^2)^2}\,,
		\label{eq:P_minimal_0}
		\\
		P^{}_1(\zeta;\mu) & =  
		\frac{2\zeta(\mu^2+\zeta^2)}{(1+\zeta^2)^3} 
		\exp\left(\frac{(1-\mu^2)\zeta^2}{1+\zeta^2}\right) \,, 
		\label{eq:P_minimal_1}
		\\
		P^{}_2(\zeta; \{\mu,\mu\}) & = 
		\frac{2\zeta}{(1+\zeta^2)^2}
		\exp\left(\frac{2(1-\mu^2)\zeta^2}{1+\zeta^2}\right)
		\frac{I_2\big(\frac{2(\mu^2+\zeta^2)}{1+\zeta^2}\big)}
		{I_0(2\mu^2)-I_1(2\mu^2)}\,.
		\label{eq:P_minimal_2}
	\ea
\end{subequations}
They are correctly normalized to 1 when integrated over 
$0\leq \zeta \leq\infty$. 
The result for $P_0$ agrees with \cite{Akemann:2008zz}.
In $P_2$ the masses were set equal for simplicity.  

A salient feature of \eqref{eq:246abc} is that they decay 
only polynomially $(\propto \zeta^{-3})$ at large $\zeta$, in contrast to 
a Gaussian decay in chGUE \cite{Forrester1993,Wilke:1997gf,Nishigaki:1998is}. 
This long tail of $P^{}_{N_f}(\zeta)$ could be a signal of weak eigenvalue repulsion 
in this model. Actually, the decay $\sim \zeta^{-3}$ can be shown 
for any $N_f$ and any masses, on the basis of \eqref{eq:Ps_rep_by_R_S}. 
If we rescale the variable as $\displaystyle x\to x/\zeta$ in the numerator of 
\eqref{eq:Ps_rep_by_R_S}, we get an additional overall factor $\zeta^{-4}$ 
while the rest of the integral tends to a well-defined large-$\zeta$ limit. 
Combined with $\zeta$ at the head of \eqref{eq:Ps_rep_by_R_S}, 
the prefactor becomes $\zeta^{-3}$.

\begin{figure}[tb]
	\begin{center}
	  	\hspace{-12pt}
	  	\mbox{
			\includegraphics[width=.5\textwidth]{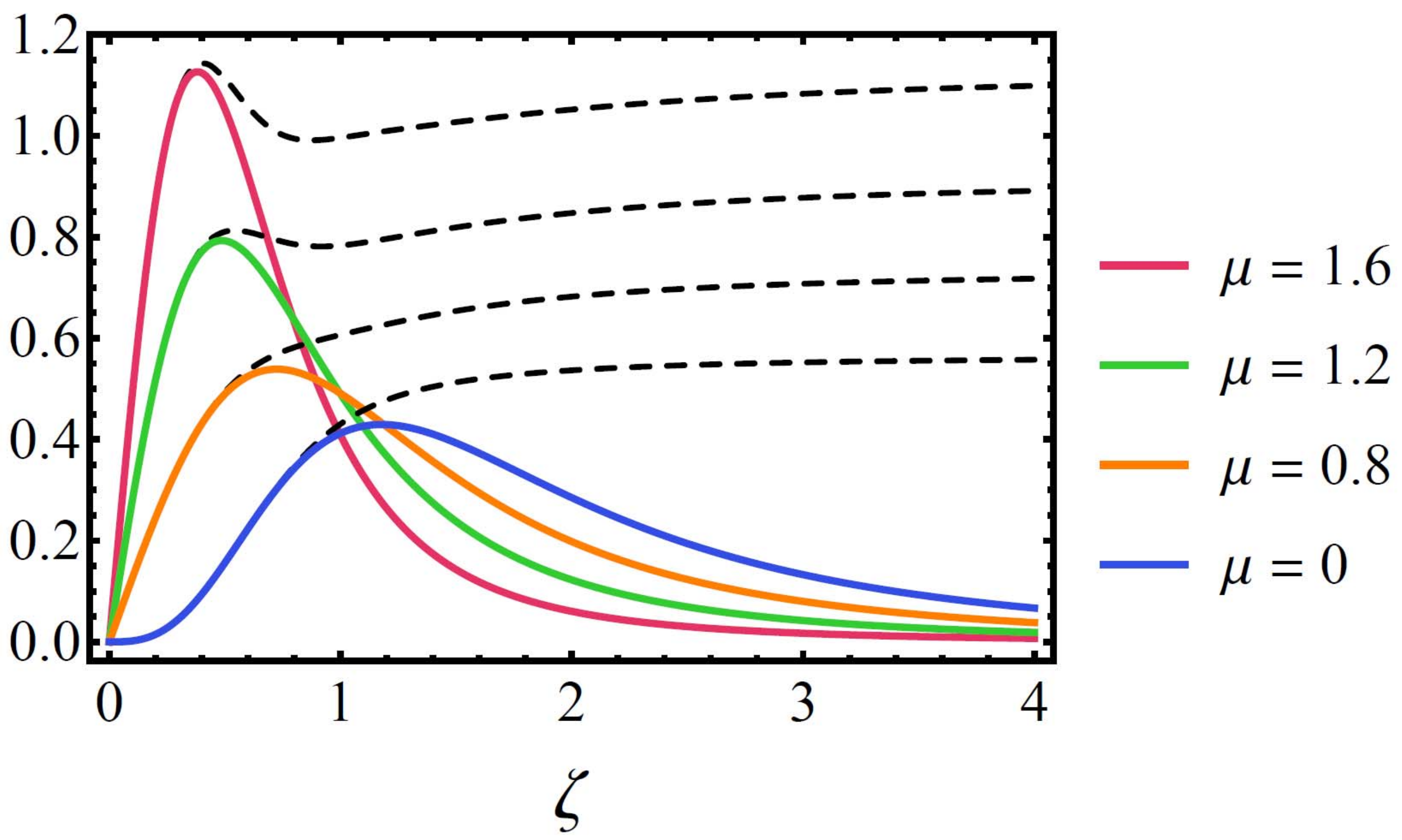}\ 
			\includegraphics[width=.5\textwidth]{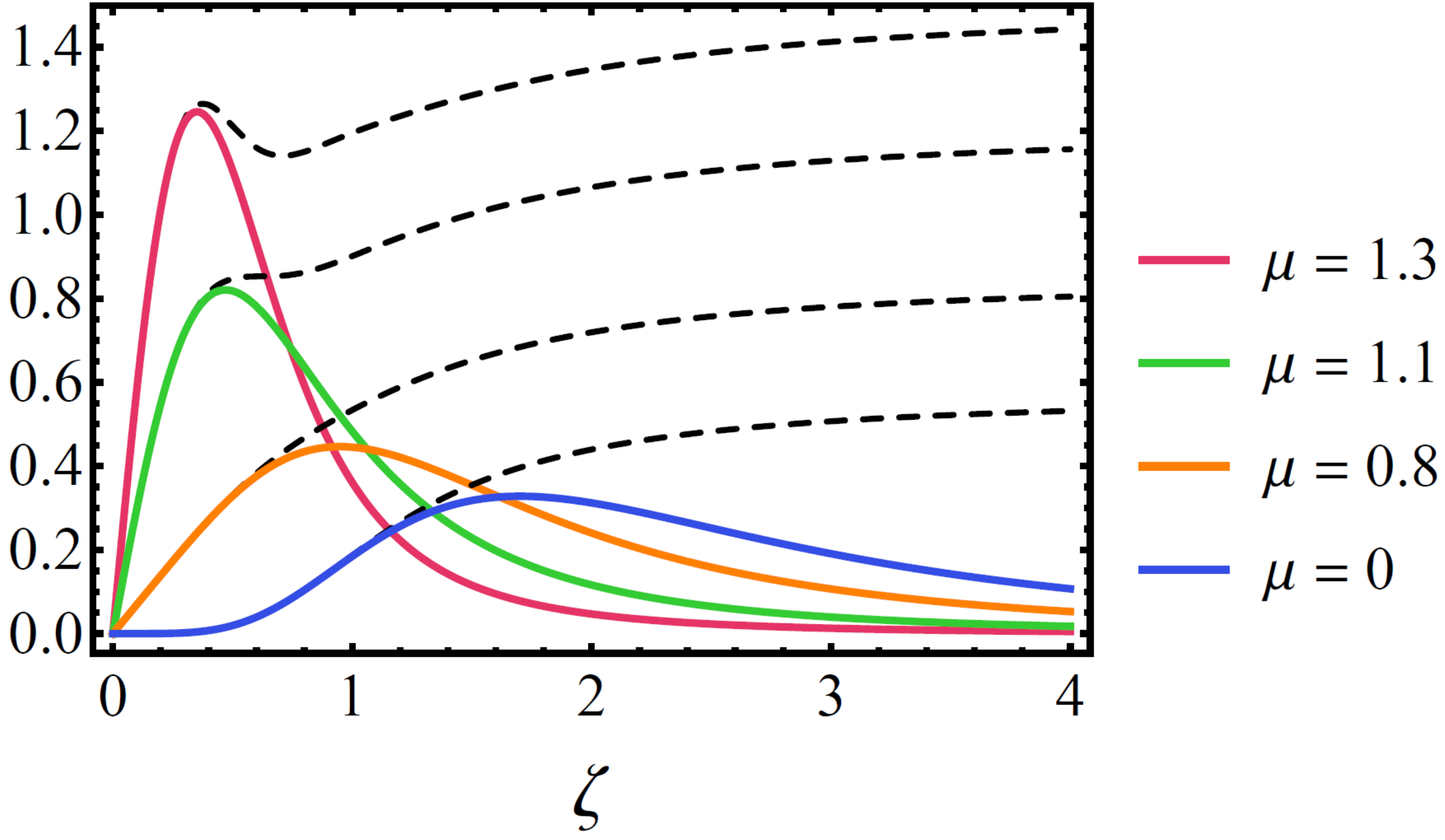}
	  	}
	  	\put(-370, 134){\large $P^{}_{1}(\zeta;\mu)$}
	  	\put(-162, 135){\large $P^{}_{2}(\zeta;\{\mu,\mu\})$}
	\end{center}
	\vspace{-1.2\baselineskip}
	\caption{Smallest eigenvalue distribution \eqref{eq:246abc} 
	for $N_f=1$ (left) and $N_f=2$ (right) for varying masses. 
	The microscopic spectral density for each mass is also shown 
	for comparison (black dashed lines). 
	}
	\label{fg:stern_minimal_eigen}
\end{figure}

In Fig.~\ref{fg:stern_minimal_eigen}, 
$P^{}_{N_f}(\zeta)$ for $N_f=1$ and $2$ are plotted 
and compared to the microscopic spectral density  
\eqref{eq:rho_massive_final}. $P^{}_{N_f}$ nicely fit  
the near-zero part of the spectral density. They tend to 
be more localized near the origin and represent a peak 
in the density when the masses are increased, as anticipated
from the reduction to chGUE discussed in Sec.~\ref{sc:largemass}.  

We expect that the extension of our analysis to the $k$-th 
smallest eigenvalue distribution for general $k\in\mathbb{N}$ 
would be straightforward along the lines of \cite{Damgaard:2000ah}.

\section{\label{sc:conc}Conclusions and outlook}

In this paper we proposed an unorthodox chiral random matrix model  
with a heavy tail. The model is a one-parameter reweighting 
of the standard chGUE and can be solved exactly. 
We discussed potential relevance of this model to the 
Stern phase of QCD, where chiral condensate is zero but 
chiral symmetry is broken by higher-order condensates. 
We analytically obtained the microscopic spectral density 
and the smallest eigenvalue distribution in the large-$N$ limit and 
discussed their dependence on the number of flavors and quark masses. 
Our model is not only useful as a conceptual toy model 
for the Stern phase but may also help a numerical evaluation of 
low-energy constants in future lattice simulations through 
fitting to the lattice Dirac spectrum. 

There remain several issues that call for further investigation. 
While we only solved the model with unitary symmetry ($\beta=2$), 
it would be technically straightforward to generalize it to $\beta=1$ and $4$; 
in fact, this has already been done for the case of $N_f=0$ in \cite{Akemann:2008zz}. 
One can also build a non-Hermitian extension of this model 
in the spirit of \cite{Osborn:2004rf} by replacing the matrix 
$X$ in \eqref{eq:rmt_stern} by a sum of two independent random matrices. 
While these extensions are intriguing at least from a mathematical 
point of view, their physical meaning remains to be clarified. 
Another unanswered question is whether the present model could be applied to 
QCD in three dimensions. To the best of our knowledge, there is so far 
no study of a Stern-like phase in $2+1$ dimensions. 
We wish to address some of these issues in future work.

\acknowledgments 
This work was supported by the RIKEN iTHES project. 
The author thanks T. Wettig and J.J.M. Verbaarschot 
for valuable comments.

\bibliography{Stern_RMT_v6.bbl}
\end{document}